# Local damage detection in rolling element bearings based on a Single Ensemble Empirical Mode Decomposition

Yaakoub Berrouche[1], Govind Vashishtha [2*], Sumika Chauhan[2], Radoslaw Zimroz[2]

[1]Faculty of Technology, Department of Electronics, Université Ferhat Abbas Sétif 1, Sétif, Algeria; Email: berrouchey@univ-setif.dz

[2]Faculty of Geoengineering, Mining and Geology, Wroclaw University of Science and Technology, Na Grobli 15, 50-421 Wroclaw, Poland, Email: govind.vashishtha@pwr.edu.pl; sumika.sumika@pwr.edu.pl; radoslaw.zimroz@pwr.edu.pl

* Corresponding author, *Govind Vashishtha: govindyudivashishtha@gmail.com; govind.vashishtha@pwr.edu.pl

**Abstract:** A Single Ensemble Empirical Mode Decomposition (SEEMD) is proposed for locating the damage in rolling element bearings. The SEEMD does not require a number of ensembles from the addition or subtraction of noise every time while processing the signals. The SEEMD requires just a single sifting process of a modified raw signal to reduce the computation time significantly. The other advantage of the SEEMD method is its success in dealing with non-Gaussian or non-stationary perturbing signals. In SEEMD, initially, a fractional Gaussian noise (FGN) is added to the raw signal to emphasize on high frequencies of the signal. Then, a convoluted white Gaussian noise is multiplied to the resulting signal which changes the spectral content of the signal which helps in extraction of the weak periodic signal. Finally, the obtained signal is decomposed by using a single sifting process. The proposed methodology is applied to the raw signals obtained from the mining industry. These signals are difficult to analyze since cyclic impulsive components are obscured by noise and other interference. Based on the results, the proposed method can effectively detect the fault where the signal of interest (SOI) has been extracted with good quality.

## 1. Introduction

Bearings are crucial parts of any industry whose health status directly affects the performance of any mechanical system. The vibration signals obtained from the bearings are complex as they contain background noise making it difficult to extract useful information. The health status of the bearings should be evaluated at regular intervals to avoid any catastrophic consequences through fault diagnosis.

Researchers have utilized some of the fault detection techniques to serve this purpose. For instance, Wang [1] introduced several indicators useful for damage detection as spectral L2/L1 norm, spectral smoothness index and spectral Gini index. Borghesani [2,3] proposed cyclo-stationary analysis with logarithmic variance stabilisation and some log-envelope indicators. Mauricio et al. [4] found that the source of impulsive behaviour might be also related to electromagnetic interference. Yu et al. [5] proposed a statistical modelling based on alpha-stable distribution for local damage detection. This approach has been also exploited by other authors [6–10].

The researchers have put forwarded time-frequency analysis (TFA) to address issues caused by the signal's nonstationary nature [11][12]. Popular examples include the short-time Fourier transform (STFT) [13], the continuous wavelet transform [14], the S-transform [15], the Wigner-Ville distribution [16], and so on. These approaches handle time-varying issues by converting 1-D time-domain signals to 2-D time-frequency domains. However, due to the Heisenberg uncertainty principle and cross-term interferences, these approaches fail to achieve ideal time and frequency resolution for signals with significant time-varying characteristics and closely spaced frequencies.

Gabor suggested that tensorisation of spectrogram may be very effective approach for damage detecion in noisy vibrations [17]. Combet and Gelman [18] utilized the spectral kurtosis as a example of optimal Wiener filter. Wodecki et al. [19] used genetic algorithm to optimise filter design procedure. Smith et al. [20] proposed optimisation scheme for demodulation band selection. Antoni [21] proposed the concept of infogram that in some sense fused information from time and frequency domain and aggregate value of negentropy to find simultaneously information from two different perspectives. It was found later by Sobkowicz et al. [22] that minor modification makes improved infogram suitable for non-Gaussian noise also. Wang [23] proposed infogram based Bayesian inference for bearings fault detection procedure improvement. Wang et al. [24] proposed novel optimal demodulation frequency band selection method called traversal index enhanced- gram (TIEgram) for rolling bearing fault diagnosis. Mauricio et al. [25] introduced a multiple criteria band selection optimisation called IESFO-gram. Kruczek et al. [26] proposed novel cyclostationary measures appropriate for non-gaussian noise. Consequently he proposed generalized spectral coherence for cyclostationary signals with *α*-stable distribution [27]. Randall sumarised current approaches for bearings diagnosis in his tutorial paper [28]

Huang et al. [29] introduced the Empirical mode decomposition (EMD) approach, which is a significant achievement in the field of signal processing, to realize the adaptive decomposition of non-stationary signals. This approach may breakdown complicated signals into a number of intrinsic mode functions (IMFs) based on the signal's local properties on the time scale. Since its introduction as a strong signal analysis tool, EMD has been widely explored and useful to fault identification. EMD, on the other hand, lacks a mathematical basis and suffers from issues like mode mixing and upper or lower envelopes.

To overcome this drawback, the researchers have put forwarded the various versions of EMD including ensemble EMD (EEMD) and complete ensemble EMD (CEEMD) methods by including white noise in it [30,31]. Park et al. [32] performed research on the diagnosis and classification of gear failures using EEMD and produced good results. Chen et al. [33] used EEMD adaptive stochastic resonance for extracting the weak fault features of planetary gear. Unfortunately, EEMD is unable to eliminate the extra Gaussian noise and mode mixing problem totally.

After that, Dragomiretskiy et al. [34] proposed the variational mode decomposition (VMD) approach, which has a solid theoretical basis and abandons the recursive screening notion used in EMD. It adapts the acquisition of signal mode components to the variational framework innovatively. The signal frequency domain decomposition and mode component separation are done by creating and solving the limited variational model. Since the Wiener filter was introduced into VMD's mode update mechanism, its noise robustness and anti-mode mixing performance have been improved [34]. VMD has recently achieved great advances in mechanical defect detection [35]. However, analyzing the wideband non-stationary signal with overlapping spectrum is not optimal. From the literature, it is clear that the famous signal decomposition techniques are associated with some drawbacks that affect their performance which necessitates the need for advanced signal decomposition techniques.

## 2. Problem formulation

Raw vibration or acoustic signals recorded for damage detection to produce the desired results. Furthermore, most damage detection algorithms are based on amplitude demodulation, often known as envelope spectrum analysis. It is generally known that for demodulation, a mono-component carrier should be considered, therefore the signal should be pre-filtered to pick the appropriate signal to demodulate while minimizing the influence of non-informative components. There are several strategies for enhancing the signal. One highly rich class is to construct filter characteristics based on statistical analysis of spectral content (the aforementioned spectral kurtosis is an excellent example). Another major approach is based on signal decomposition. Wavelet and EMD-based methods are the most common here. Both approaches have merits and weaknesses, however, it should be noted that EMD is a data-driven technique that does not need knowledge of the fundamental wavelet form or decomposition level. So, in this study, we describe a research objective as decomposing raw vibration/acoustic signals to increase SNR and extracting Signals of Interest (SOI) containing cyclic and impulsive signals connected to the damaged bearing part. Let us assume that the observed signal is an additive combination of SOI and noise:

$$s(t) = SOI(t) + n(t) \tag{1}$$

The SOI is a cyclic and impulsive signal with the period $T$ related to the expected fault frequency. The $n(t)$ is the random component that could be a Gaussian or non-Gaussian distributed noise. For simplicity, we assume that $s(t)$ doesn't contain another deterministic component that has been removed from observation by any random/deterministic components separation technique. It should be mentioned that it does not affect the procedure, just simplifies the model of the signal.

## 3. Single Ensemble Empirical Mode Decomposition

### 3.1. Classical EEMD

By using a sifting process, the EMD method decomposes an input signal $s(t)$ into a finite number of IMFs (intrinsic mode functions) components and a low-order polynomial represented by $w(t)$:

$$s(t) = \sum_{k=1}^{N} IMF_k(t) + w(t) \tag{2}$$

The main issue with EMD is a mode mixing issue for linear and stationary input signals. So, to overcome this drawback, the EEMD is introduced. The EEMD method is a noise-assisted data that adds a finite quantity of white noise to an input signal. For a set number of ensemble trials (Ne), every time, the sifting algorithm decomposes the obtained signal into IMFs. The effective ensemble of IMFs is defined as the mean of all ensemble trials produced from IMFs of the same order.

The fundamental difficulty with the EEMD approach is that before signal processing, we must choose the right amplitude of white noise and a number of ensemble trials (Ne) randomly without any defined procedure. Similarly, a large number of ensemble trials are required to neutralize the effect of additional white noise, which increases exponentially the computation time and complexity. So, the EEMD method is not suitable for real-time application such as local damage detection.

### 3.2. Fractional Gaussian noise

The fractional Gaussian Noise (FGN) is a Gaussian-centered random process that is created regularly by sampling the fractional Brownian motion phenomena $(B_H(t))$. The FGN has a higher frequency content, making it suited for dealing with high-frequency non-Gaussian noise in the signal. As a consequence, the desired signal may be simply extracted. The method for computing the first difference is described by Eq. (3):

$$s(n) = B_H(n) - B_H(n-1) \tag{3}$$

The autocorrelation FGN of lag k is given by Eq. (4):

$$R_x(m) = \frac{\sigma^2}{2}(|k-1|^{2H} - 2|k|^{2H} + |k+1|^{2H}) \tag{4}$$

Where, two parameters describe the FGN (Hurst exponent H and its variance $\sigma^2$).

### 3.3. Convolution

Convolution indicates the degree of overlap of one signal as it is moved across another signal. It is an integral operation on two signals that yields a third signal. Convolution can be thought of as a measurement of how one signal affects another. The output of the convolution of two continuous-time signals is provided by Eq. (5).

$$x(t) * h(t) = \int x(\tau) h(t-\tau) d\tau \tag{5}$$

## 4. Proposed procedure for fault detection

The SEEMD algorithm also uses the partial approach as adopted by the EEMD algorithm. However, there are two main differences. firstly, in the SEEMD method, the white Gaussian noise is substituted with the fractional Gaussian noise (FGN). Adding FGN as an excitation increases high frequencies that may highlight medium and high resonances and decreases low frequencies that contain non-informative high amplitude components. It has these properties: The power spectral density (PSD) increases with the frequency. It means that we have a high PSD for the high frequency [36]. It is noted that at higher frequencies and bigger Hurst indexes (H=0.9), the power spectral density (PSD) of FGN evolves either extremely slowly or flatly with frequency. Thus, FGN may be roughly considered a wide-band noise in the high-frequency range. The FGN can handle high-frequency noise in the signal because it has greater high-frequency richness. So, FGN emphasises some frequencies (high frequencies) of the input signal [37]. So, the spectral characteristic will be changed from flat (WGN) to high-frequency excitation which is suitable to extract a weak periodic signal (SOI). Likewise, we have an extra operation before the sifting process.

In the SEEMD algorithm, the raw signal is firstly added to FGN with Hurst exponent H equal 0.1 where the FGN is considered as negatively correlated, antiperspirant, and the process is short-term dependent. That means that the power spectral density (PSD) increases with the frequency [38]. The addition of high-frequency signals improves the performance of the weak periodic signals [39].

The next step is to multiply the resulting signal by a convoluted white Gaussian Noise (WGN). A Convoluted WGN (WGN*WGN) affects the signal in a non-uniform manner across all frequencies. Not like just WGN which is flat in the spectral domain and affects all frequencies of the signal by the same noise level. Here, we have computed the Convolution of WGN to change the spectral characteristic of WGN. So, It causes changes in the signal's spectral content or behaviour from flat to low medium frequency excitation which is suitable for extracting a weak periodic signal (SOI). When noise is convoluted, it introduces a frequency-dependent effect. It indicates that the noise's impact on a signal is not uniform across all frequencies. The convoluted WGN noise means that certain frequency components are emphasized (have more energy) or attenuated. The convoluted WGN, compared to WGN, has more energy at lower frequencies compared to higher frequencies as shown in Fig. 1.

When we apply it to the resulting signal in a nonlinear manner (multiplication), the specific idea in the SEEMD method, it causes changes in the signal's spectral content or behavior. So, the high-frequency excitation added in a nonlinear manner helps the extraction of weak periodic signals.

Finally, the obtained signal is decomposed by using a sifting process in order to obtain all IMFs and a residue.

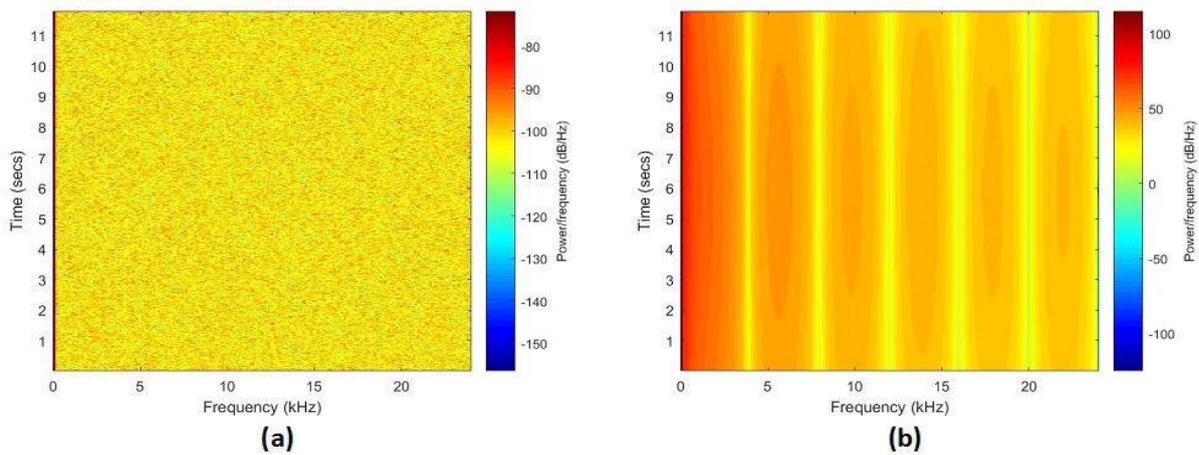

**Fig. 1** Spectrogram of **(a)** WGN **(b)** convoluted WGN

The proposed methodology for fault identification is shown in the Fig. 2.

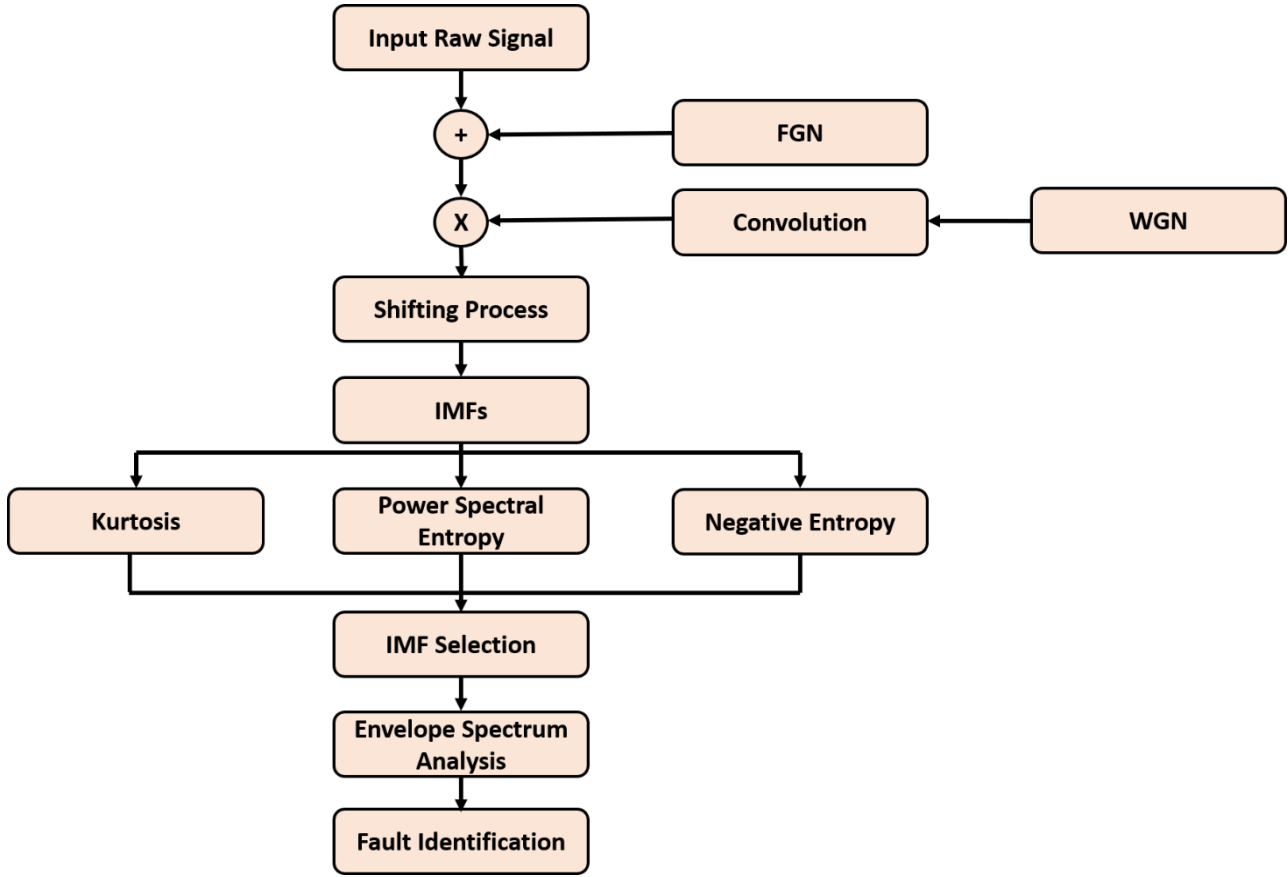

Fig. 2 Procedure for fault detection

## 5. Simulated data analysis

The proposed method has been applied to a simulated signal representing a defect on the ball of the ball bearing whose detailed explanation is given in Ref. [40]. Some important parameters of the simulated signal are tabulated in Table 1.

Table 2
Parameters for simulated signal

| Description | Value |
| --- | --- |
| Bearing roller diameter (d) in mm | 8.4 |
| Pitch circle diameter (D) in mm | 71.5 |
| Contact angle [rad] | 15.7*pi/180 |
| Number of rolling elements (n) | 16 |
| Fault Type | Ball defect |
| Sampling frequency (fs) | 20000 |
| Rotation frequency profile (fr) | fc+*pi*fd.*(cumsum(cos(fm.*theta)/N)) |
| Carrier frequency (fc) | 20 |
| Modulation frequency (fm) | 0.1*fc |
| Frequency deviation (fd) | 0 |
| Number of points per revolution (N) | 500 |
| Signal to noise ratio [dB] | 20 |
| Amplitude modulation at the fault frequency (qFault) | 10 |
| Amplitude value of the deterministic (qStiffness) | 0.1 |
| Amplitude value of the deterministic component related to the bearing rotation (qRotation) | 0.1 |

The corresponding raw signal and spectrogram is shown in Fig. 3. It can be observed from the figure that the modulating frequency is submerged with the noise. It is clear from the Fig. 3(b), that there is no definite informative frequency band is visible making it difficult to extract the fault features.

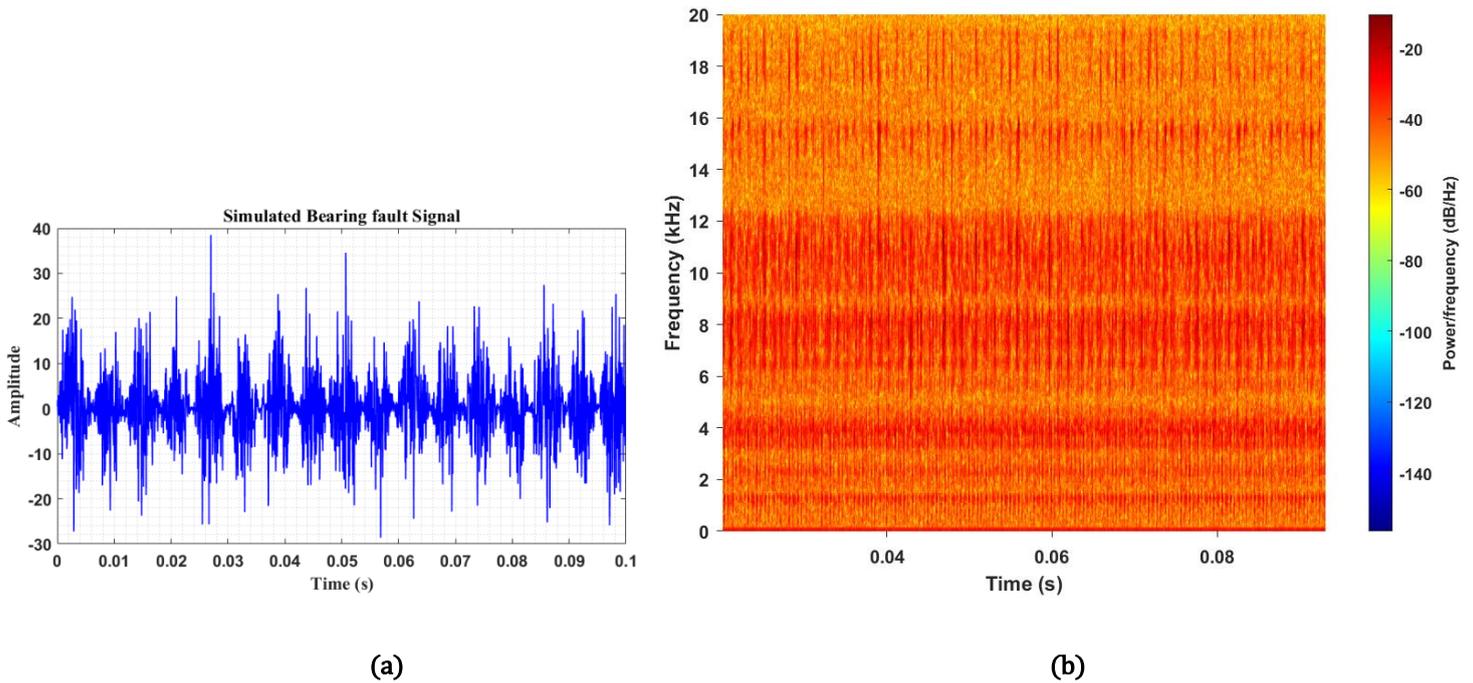

(a)                                                                 (b)

**Fig. 3** Simulated bearing signal for ball defect

The simulated signal is treated with the proposed methodology which decompose the raw signal into different IMFs as shown in Fig. 4. A total of 11 IMFs have been obtained. Out of which, IMF 3 is the one which showing higest impulsive. The kurtosis value for the raw signal is found to be 4.8985 whereas the kurtosis value for the IMF 3 is found to be 5.9568. The prominent informative frequency band is visible in spectrogram of the IMF 3 as shown in Fig. 5 (b). The superiority of the proposed methodology is validated on the simulation analysis. To further investigate the efficiency, the proposed method has been applied to the real data analysis in the subsequent sections.

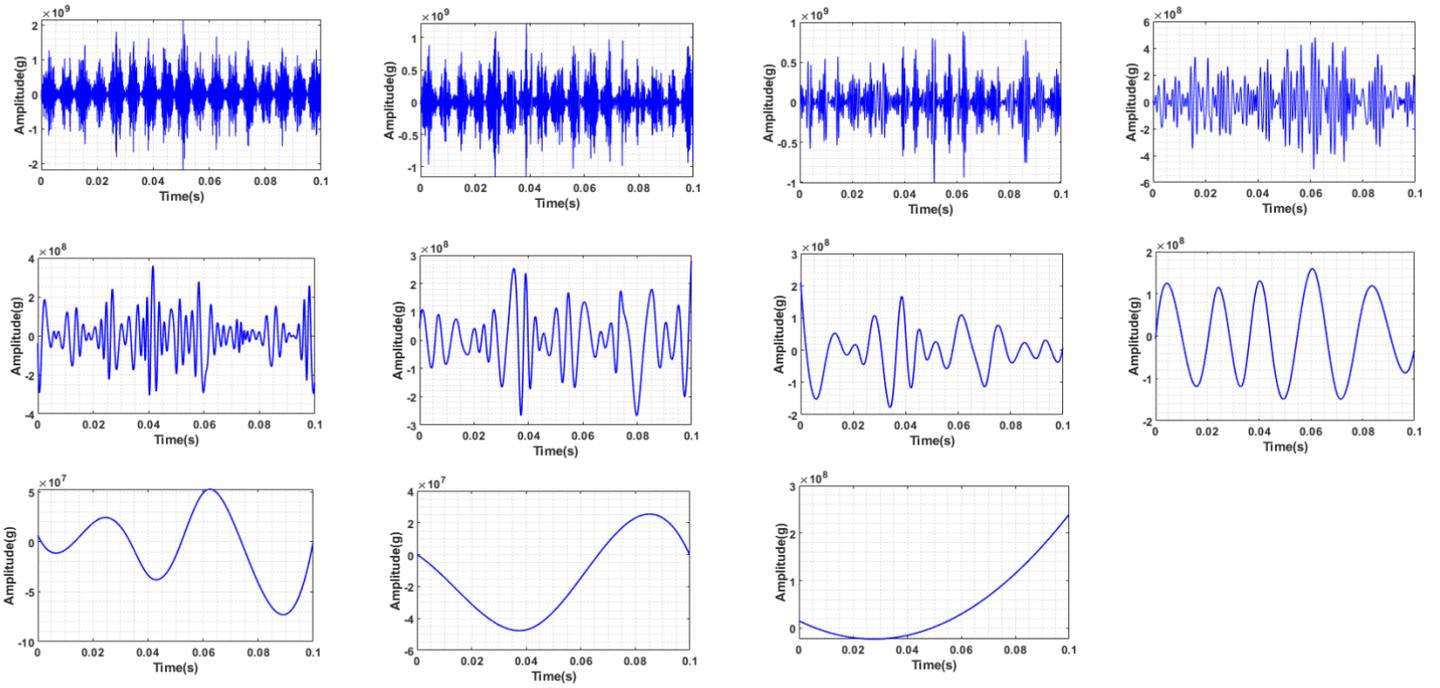

Fig. 4 Decomposition of simulated signal through SEEMD

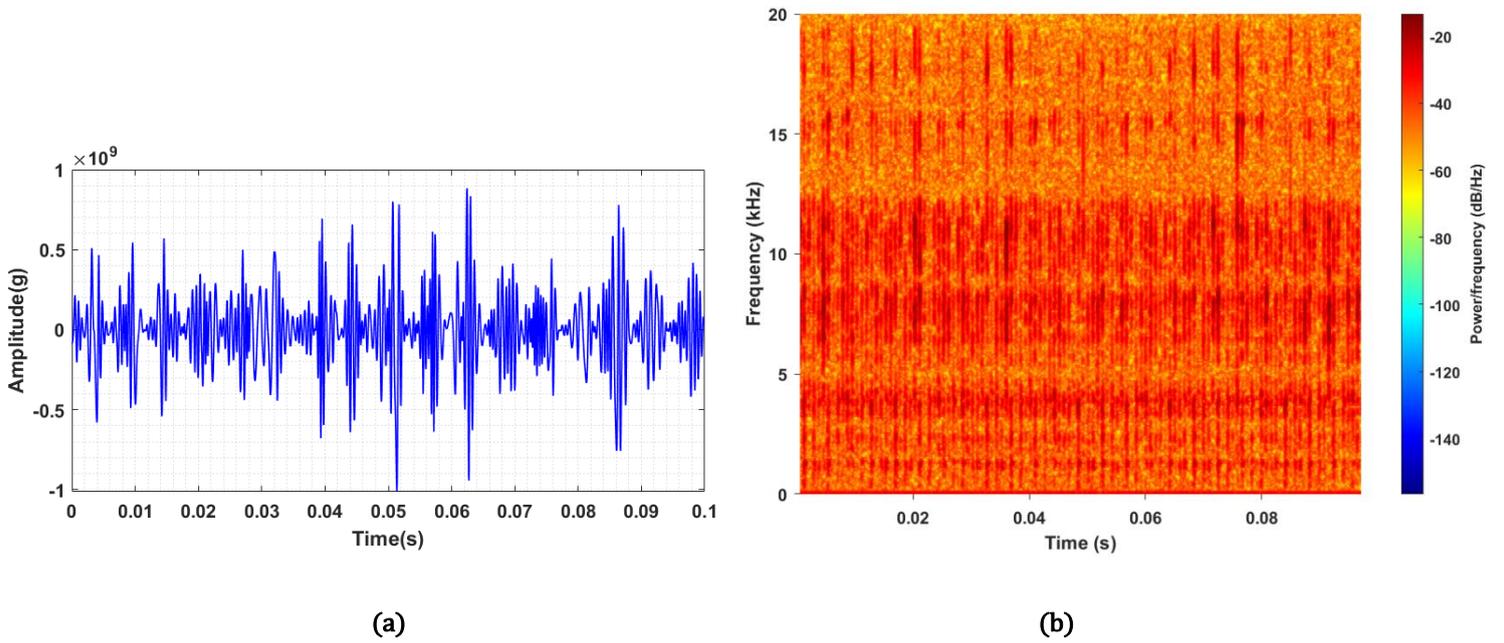

(a)                  (b)

Fig. 5 Result of SEEMD-based decomposition of simulated signal for ball defect

6. Real data analysis

High-power machines with complicated construction and time-varying operating conditions are employed in the mining sectors. The conveyor driving station is made up of several drives with capacities ranging from 630 to 1000 kW. In our situation, two 1000 kW drives were employed. As illustrated in Fig. 6, the drive unit consists of an electric motor, a connection, and a two-stage gearbox coupled with a pulley. The pulley is attached to the shaft and balanced by the two bearing sets. The pulley is rubber-coated to improve friction between the pulley and the belt. The stiff coupling was used to make the connection between the pulley and the gearbox. The belt conveyor systems have produced both vibration and sound signals. To validate its efficacy, the suggested methodology was used for vibration data from the belt-conveyor system acoustic data from idlers.

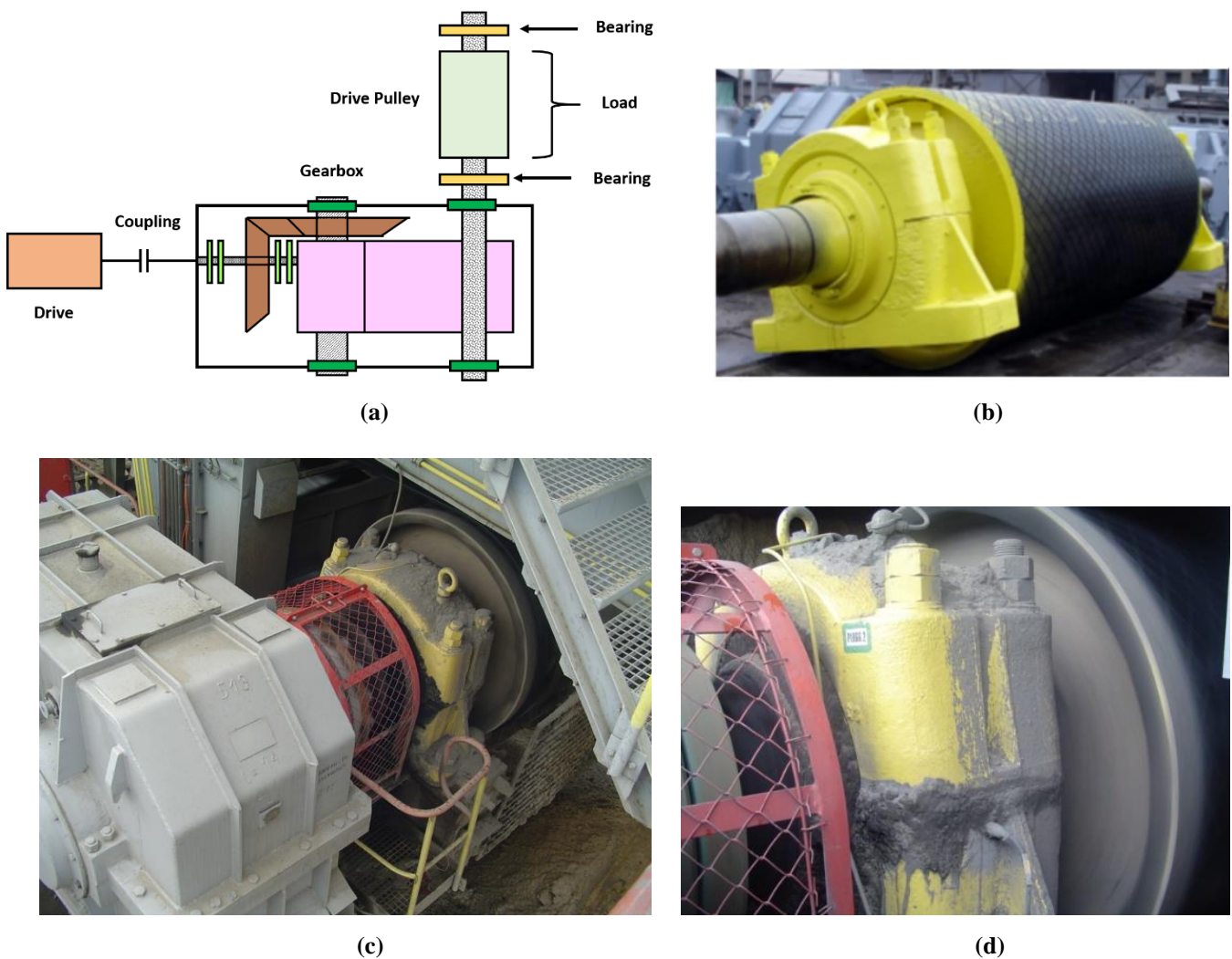

**Fig. 6. (a)** Drive unit for belt conveyor **(b)** pulley with bearing housing mounted on the shaft **(c)** view on the coupling between gearbox and pulley, and **(d)** view on sensor location on the pulley

## 6.1. Analysis of the vibration signal from rolling element bearing in a pulley

The vibration signals have been recorded from the bearings that support the pulley as presented in Fig. 6 (c). The accelerometer is fixed to the bearing hub through a screw as shown in Fig. 6 (d). The sampling frequency is set to 19.2 kHz while acquiring the vibration signals. The fault frequency is found to be at 12.6 Hz.

The waveform in time and time-frequency domain of the defective bearing of the belt conveyor system are shown in Fig. 7 (a) & (b). The value of kurtosis obtained in the case input signal is 3.0978 which is quite low due to the interferences which makes it difficult to extract sensitive information about impulsiveness from the signal with this low value of kurtosis. Also, the strong noise and other interferences generated by other components are masked over the informative signals as observed from the spectrogram making it obvious that identification of the fault features is difficult. The corresponding envelope spectrum also does not give information about the fault characteristic frequency associated with the bearing defect as depicted in Fig. 7 (c).

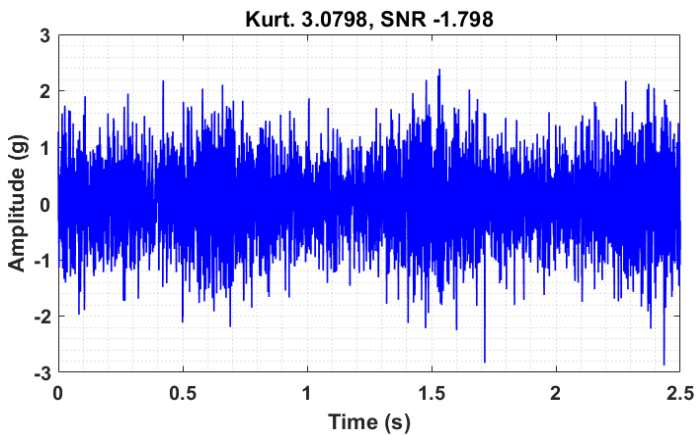

(b)

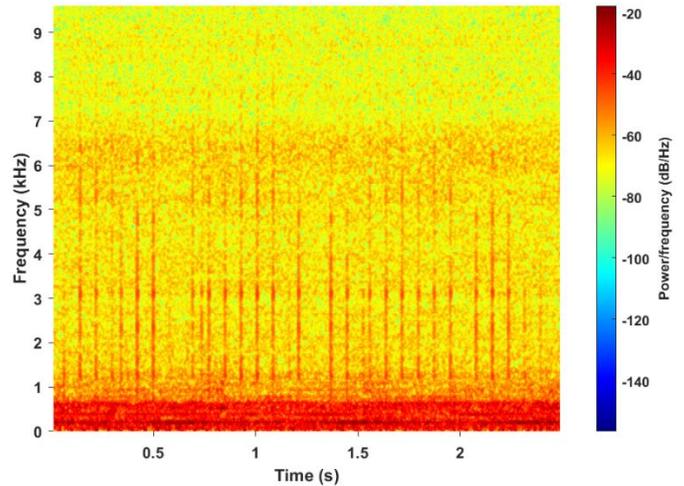

(b)

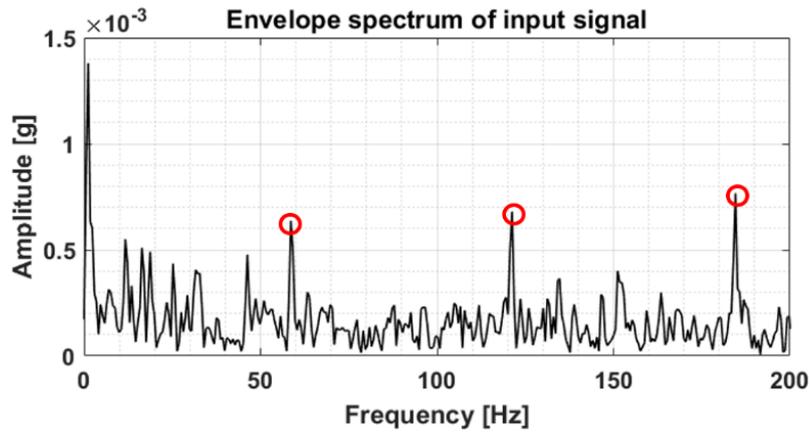

(c)

**Fig. 7 (a)** Raw vibration signal from bearing used in pulley **(b)** its spectrogram and **(c)** envelope spectrum

The proposed SEEMD has been applied to the raw bearing signal from the pulley. The resulting decomposed signals (IMFs) are shown in Fig. 8. A total of 16 IMFs have been obtained by the proposed SEEMD. Out of 16 IMFs, IMF 1 represents the most impulsive signal whereas IMF 16 represents the residual signal.

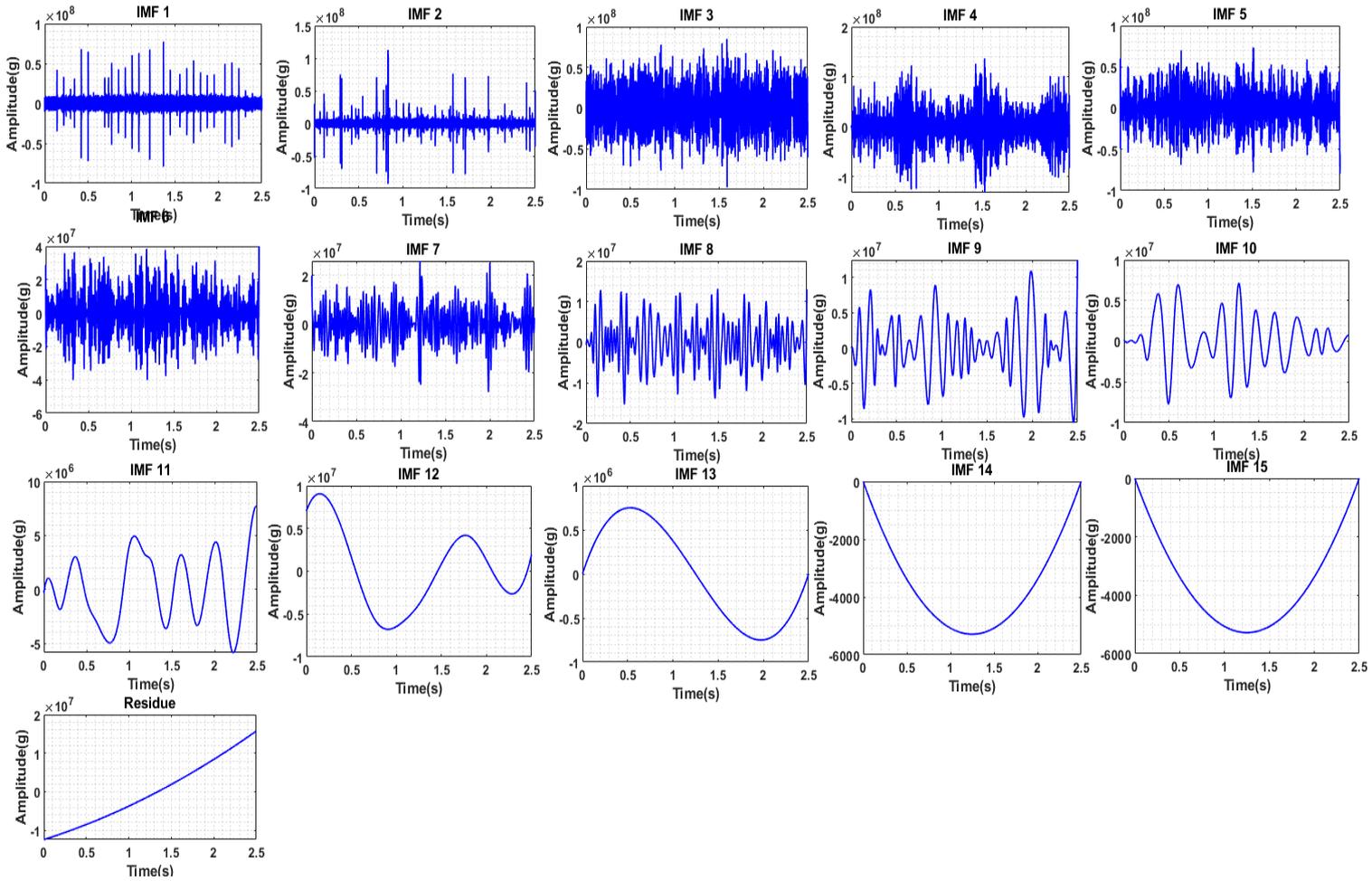

**Fig. 8** Decomposition of bearing vibration signal through SEEMD

The value of the kurtosis of the most impulsive signal i.e. IMF 1 has been tremendously increased to 96.5438. The spectrogram of the IMF 1 has also been obtained as shown in Fig. 9 (b). It is clear from Fig. 9 (b) that the proper informative frequency band is observable in the IMF 1 which proves the superiority of the SEEMD. Also, in the case of the envelope signal, it can be observed that SEEMD has successfully suppressed the energetic low frequencies resulting in high impulsiveness.

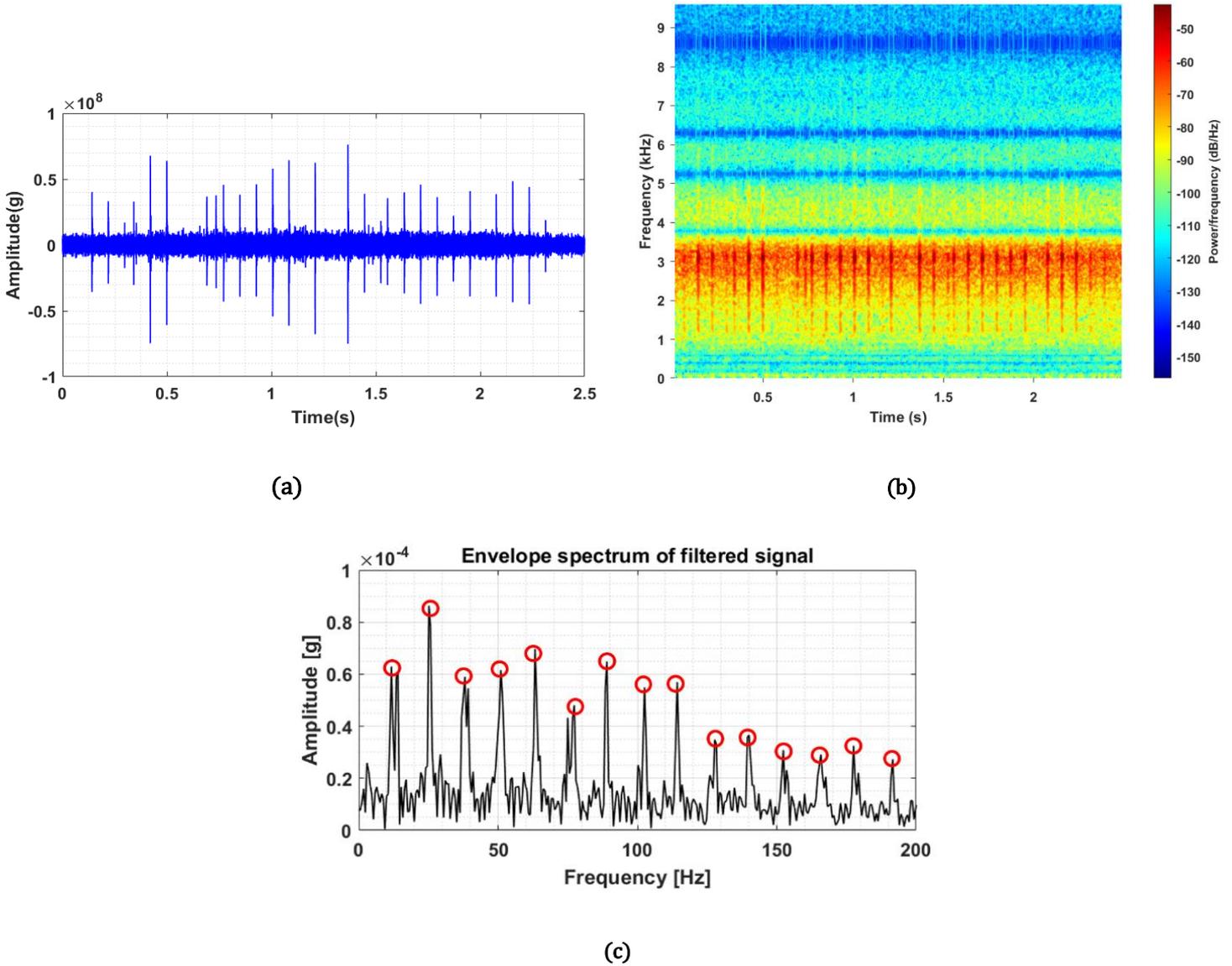

(a)　　　　　　　　　　　　　　　　　　(b)

(c)

**Fig. 9** Result of SEEMD-based decomposition of vibration signal of bearing from pulley

## 6.2. Analysis of the acoustic signal from the rolling element bearing used in the idler

The second scenario studied involves acoustic signals from the conveyor belt idlers. There, we investigate audio recordings taken using a mobile device (as shown in Fig. 10). The input measurement comprises a 10 second slice of an acoustic sound sampled at 48 kHz. The bearing utilized in these idlers is SKF 6204-2RSH, with a probable fault frequency of 5.5 Hz with low, slow speed variations between 105 and 110 revolutions per minute (the speed might be deemed constant locally). The mechanical belt joints (a metallic clamp used to link two parts of the belt) are the source of the non-Gaussian behaviour in the acoustic signal. During the activity, the belt moves on rolling idlers, and when the mechanical clamp comes into contact with a certain idler, an impulsive sound is produced (metallic clamp hits metallic coating of idler). Again, we examine situations without and with periodicity; however, the latter is not purposefully constructed, as one of the signals was obtained from broken equipment. The non-Gaussian impulses are mostly visible in the range from about the 9th to the 10th second, encompassing the whole frequency bands. More information about the experimental setup can be found in [41][42]. The raw signal and its corresponding spectrogram and envelope spectrum are shown in Fig. 11. The raw signal is highly affected by environmental noises and other interferences making it difficult to extract the repetitive impulses. The informative frequency bands are also not clearly visible in the spectrogram. The corresponding envelope spectrum also does not give information about the fault characteristic frequency associated with the bearing defect.

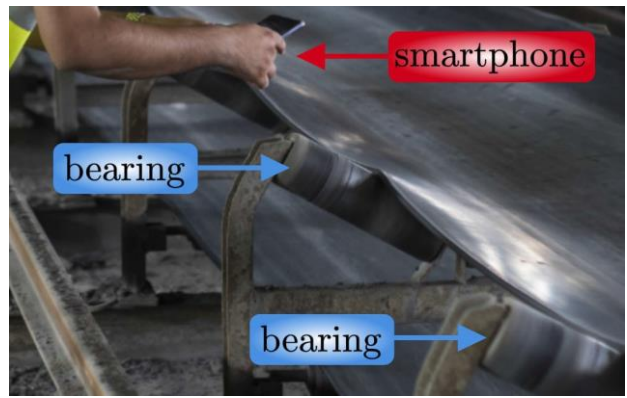

Fig. 10 Sound measurement from bearings installed in idlers.

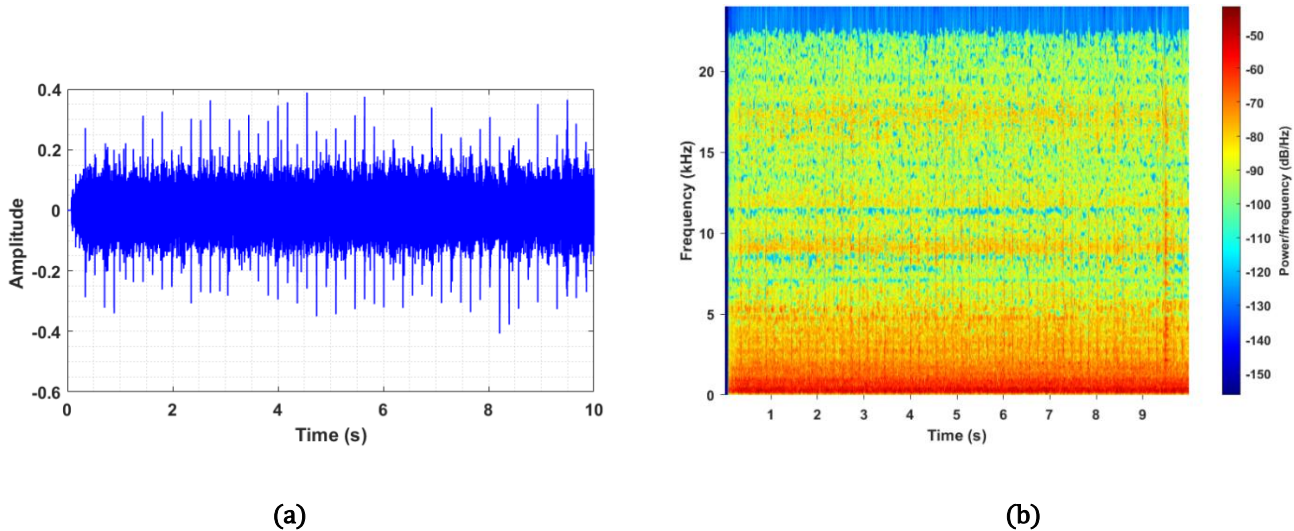

(a) (b)

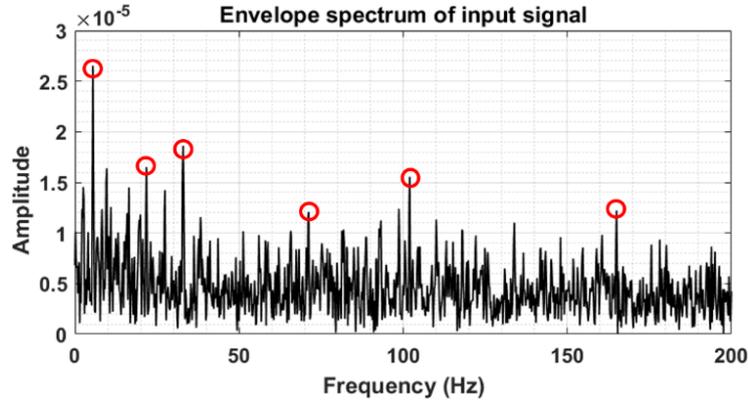

(c)

**Fig. 11 (a)** Raw acoustic signal from bearing used in idler, **(b)** its spectrogram and **(c)** envelope spectrum

The raw acoustic signal is processed through SEEMD which decomposes the signal into 19 IMFs. Out of 19 IMFs, IMF 1 is the most impulsive signal as shown in Fig. 12. One can easily observe the damage-related impulses having a kurtosis value of 95.3508, which represents an improvement of 2487.25% as shown in Fig. 13 (a). The improvement ratio of nearly 26 times the original value is the first indication of the local damage. The time-frequency structure of the input and filtered signal are represented by the spectrograms to evaluate the obtained informative frequency band as shown in Fig. 13 (b). The envelope spectrum of output signal is shown in Fig. 13(c) respectively. It can be seen from the envelope spectrum, the SEEMD is better able to suppress the energetic low frequencies exposing the impulsiveness.

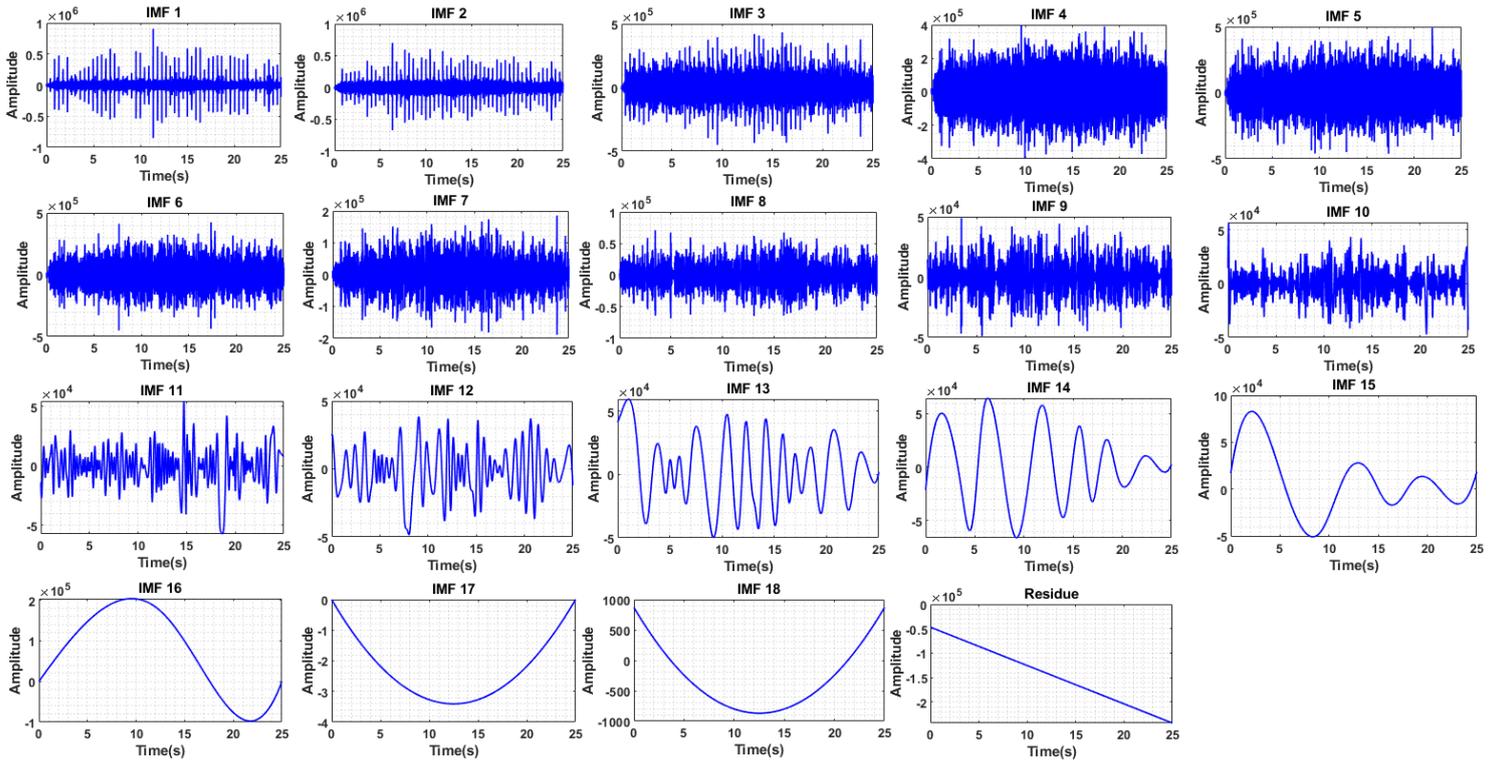

**Fig. 12** Decomposition of acoustic signal through SEEMD

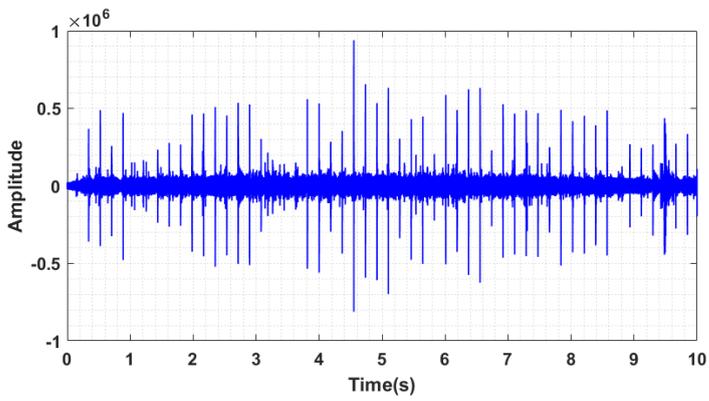
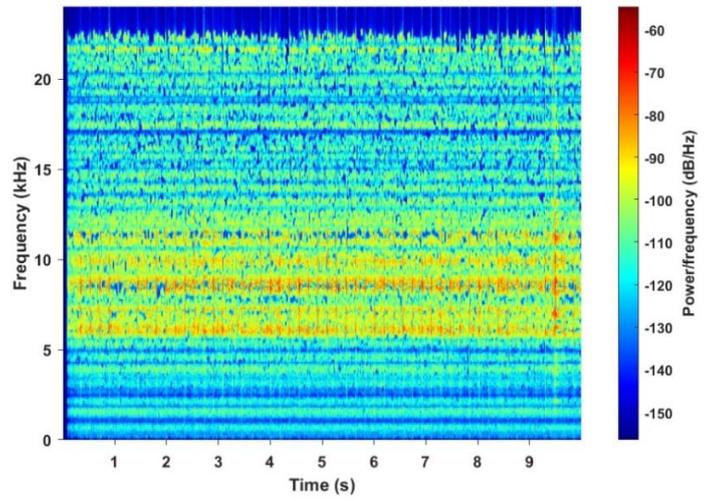

(a)  (b)

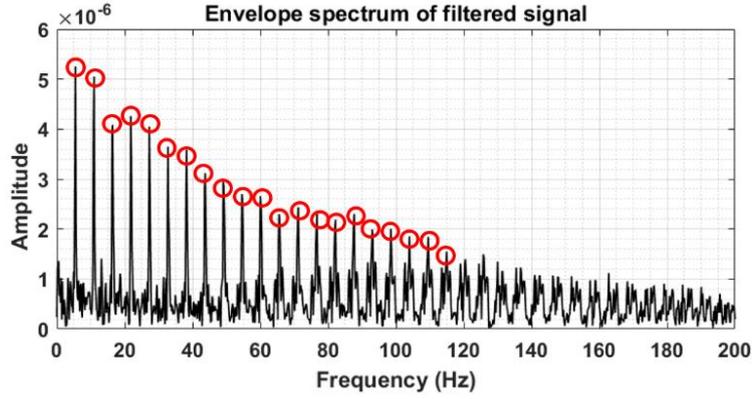

(c)

**Fig. 13** Result of SEEMD-based decomposition of acoustic signal of faulty bearing used in idler

## 7. Comparison with other methods

The proposed SEEMD is compared with existing methods to validate its effectiveness. The existing methods include EEMD and VMD.

### 7.1. Comparison with EEMD

Initially, the raw signals from the defective bearing from the pulley and the acoustic signal of the defective bearing from the idler are processed through EEMD. The decomposed signals in the form of IMFs both for vibration signal and acoustic signal are shown in Fig. 14 and Fig. 16 respectively. In the case of the vibration signal, a total of 16 IMFs have been obtained, out of which IMF 3 shows some impulsiveness. It is obvious that SEEMD performed better while extracting the impulsiveness. Also, the value of the kurtosis of the most impulsive signal (IMF 3) obtained by EEMD as shown in Fig. 15 is much less than the kurtosis value obtained by the SEEMD. The informative frequency band obtained by the EEMD is not clear enough to give sensitive information. In the case of the envelope spectrum, fault frequencies are identified are comparable to SEEMD as given in Fig. 15.

To validate the efficacy of the proposed SEEMD, the authors have utilized the Envelope Spectrum based Indicator (ENVSI) [22]. This simple indicator is used to measure the energy which is defined as:

$$ENVSI = \frac{\sum_{i=1}^{M_1} AIS(i)^2}{\sum_{k=1}^{M_2} SES(k)} \qquad (6)$$

where AIS represents the normalized amplitudes of the information signal, SES is the squared normalized envelope spectrum, $M_1$ is the number of harmonics, and $M_2$ is the number of frequency bins used to calculate the total energy.

In the case of the defective bearing from the pulley, the value of the ENVSI obtained by the SEEMD is 0.49 whereas the EEMD gives the value of ENVSI of 0.29 only.

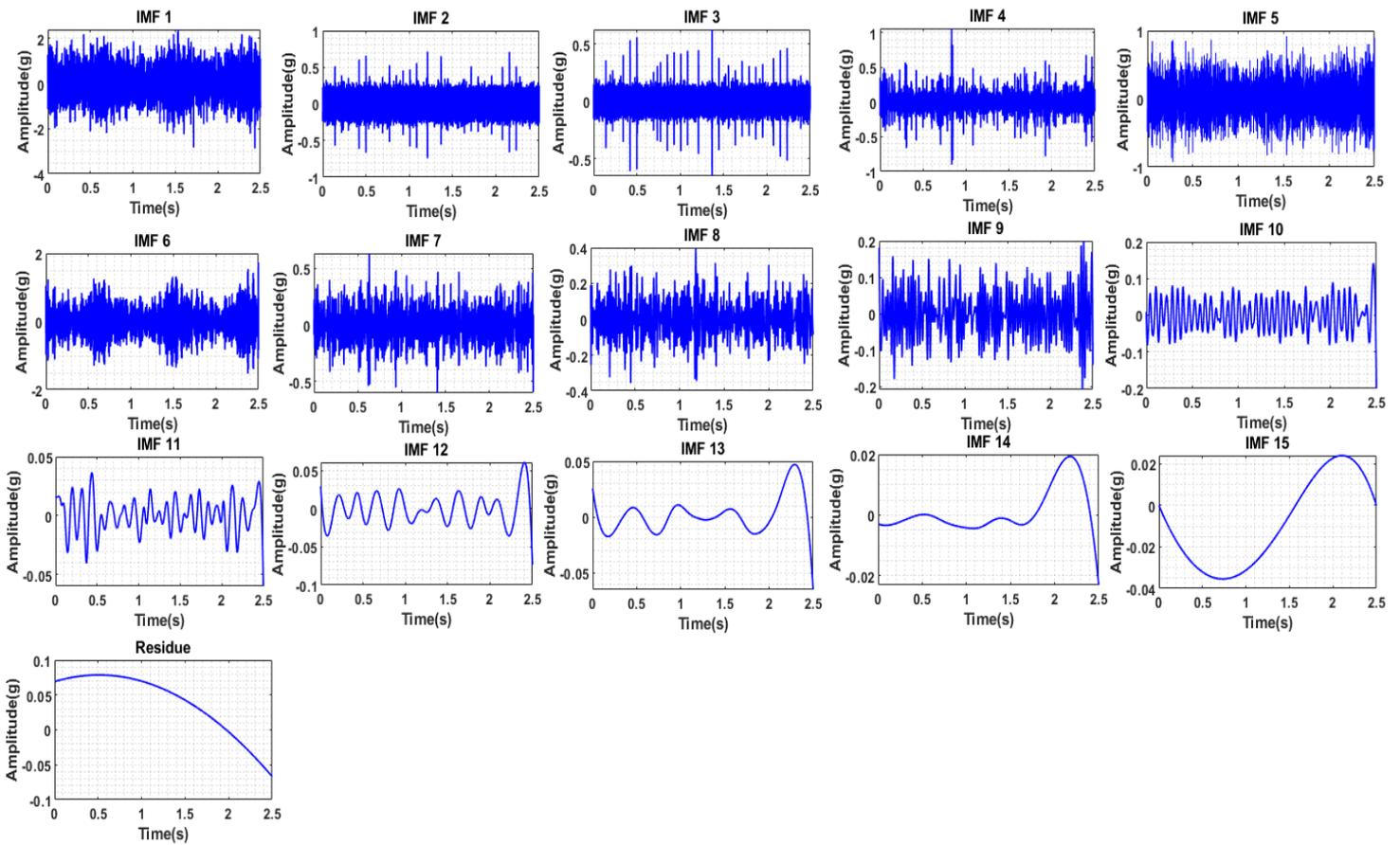

**Fig. 14** Decomposition of vibration signal of bearing from pulley through EEMD

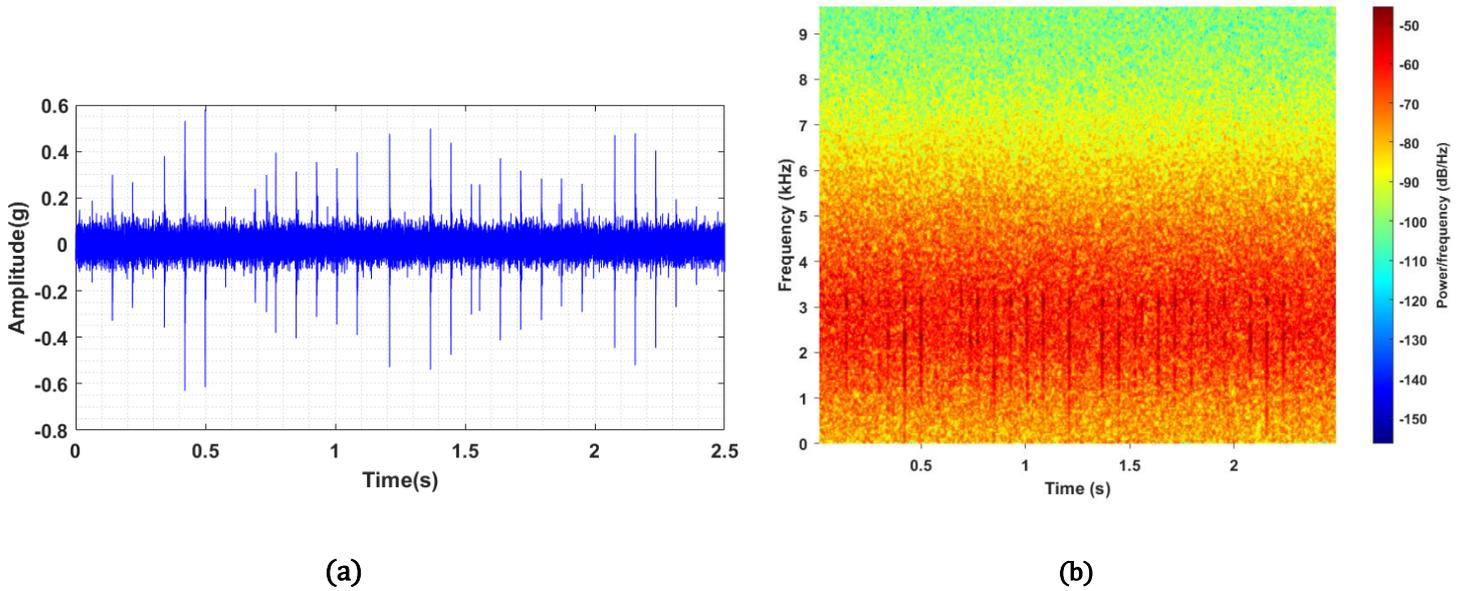

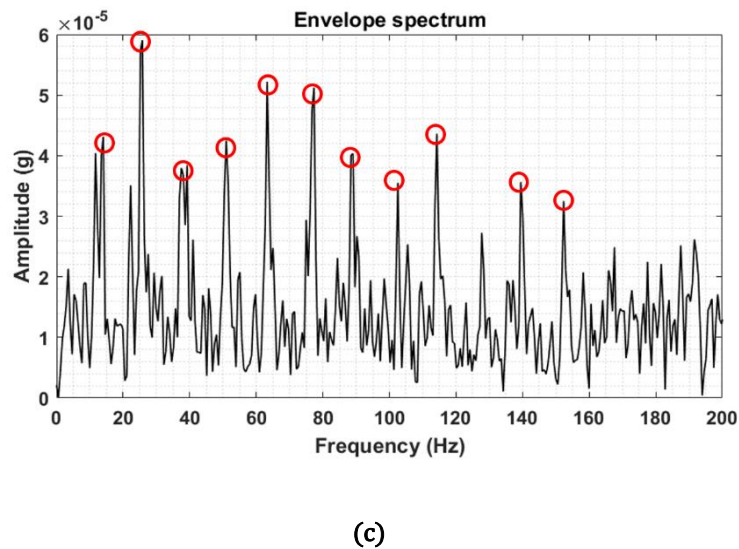

(c)

Fig. 15 Result of EEMD-based decomposition of vibration signal from faulty bearing from pulley

Whereas in the case of the acoustic signal of the defective bearing from the idler, a total of 19 IMFs have been obtained by the EEMD as shown in Fig. 16. Here the IMF 2 is the most impulsive signal. However, the kurtosis value obtained by the EEMD is much less than the SEEMD. Also, the spectrogram of the IMF 2 has not been proved effective as it fails to extract the informative frequency band as shown in Fig. 17. In the case of the envelope spectrum, fault frequencies are identified are comparable to SEEMD as given in Fig. 17 (c). The value of the ENVSI obtained by the SEEMD is 0.56 whereas the EEMD gives the value of ENVSI of 0.34 only in case of the defective bearing from the idler.

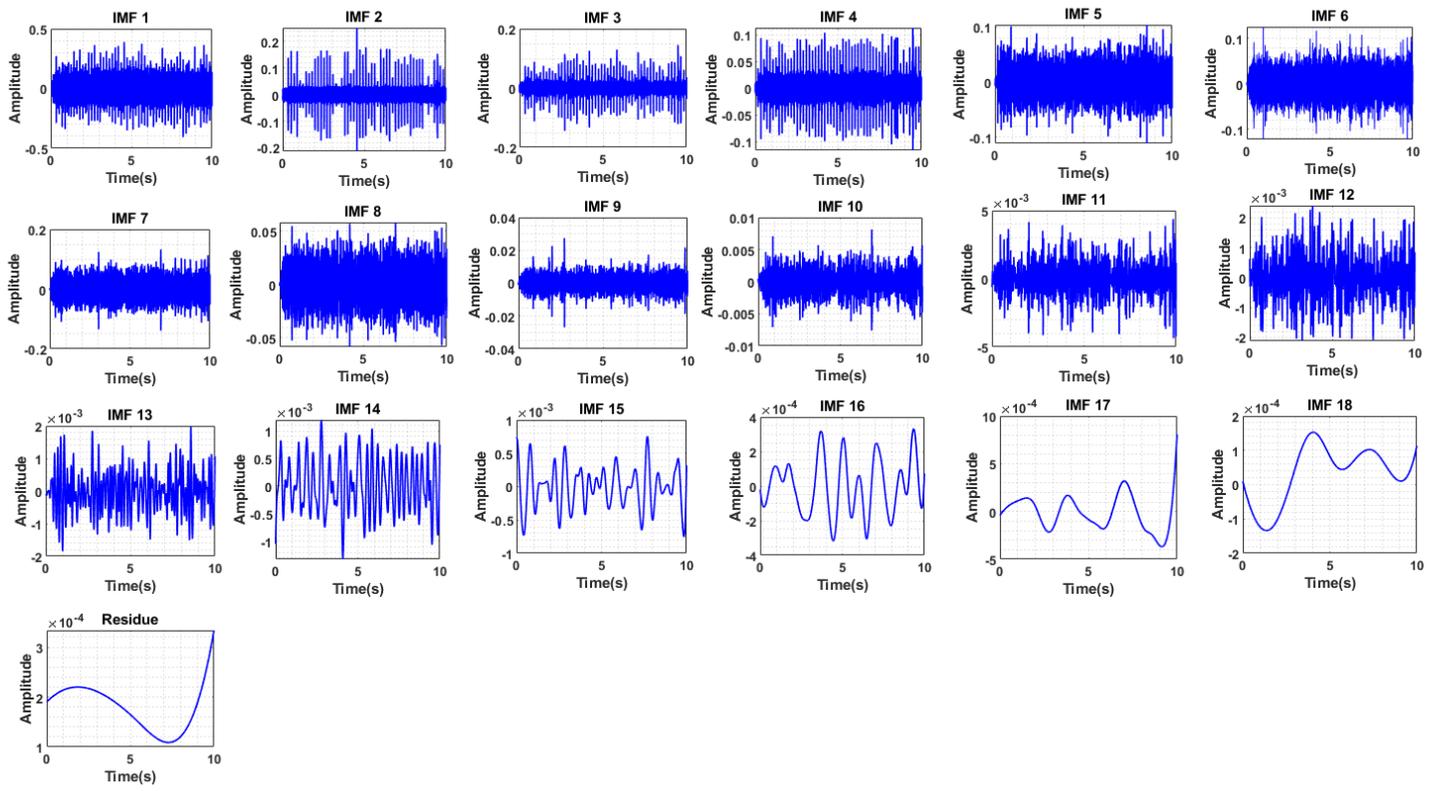

**Fig. 16** Decomposition of acoustic signal of faulty bearing from idler through EEMD

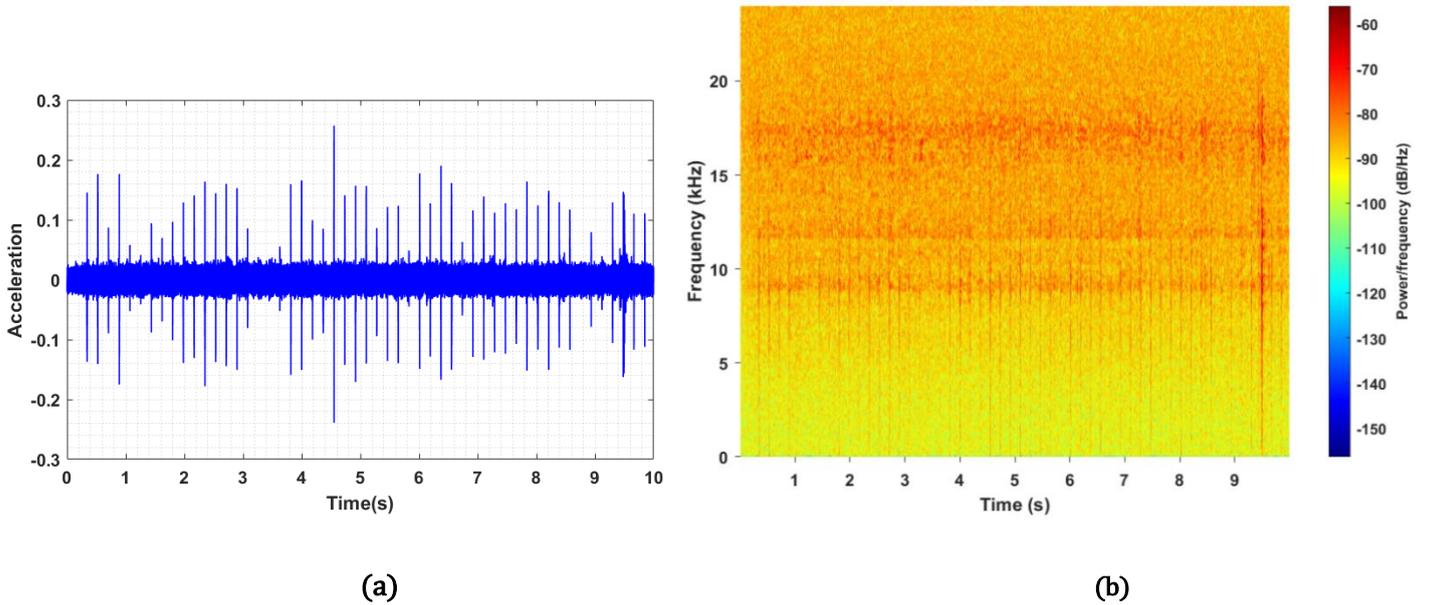

(a)  (b)

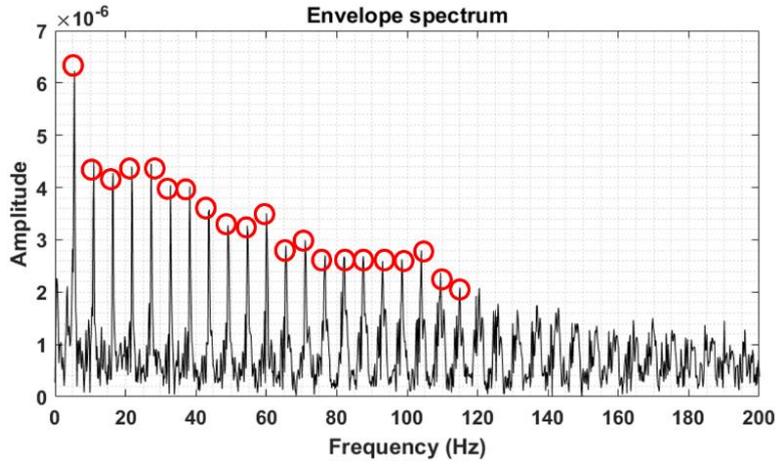

(c)

Fig. 17 Result of EEMD-based decomposition of acoustic signal from faulty bearing used in idler

## 7.2. Comparison with VMD

The VMD has also been taken into consideration for comparison. Both the signals from the defective bearing from the pulley and idler have been passed through VMD. The corresponding modes are shown in Fig. 18 and Fig. 20 respectively.

In case of the vibration signal of the defective bearing from the pulley, 8 modes have been obtained by the VMD. The best mode obtained by the VMD is mode 4 as shown in Fig. 19. The value of kurtosis is higher than EEMD but lower than SEEMD. The informative frequency band is observable in the spectrogram obtained by the VMD as shown in Fig. 19. However, the informative frequency band is highly contrasted in the spectrogram obtained by the SEEMD. In the case of the envelope spectrum, fault frequencies are identified are comparable to SEEMD as given in Fig. 19 (c).

In the case of the defective bearing from the pulley, the value of the ENVSI obtained by the SEEMD is 0.49 whereas the VMD gives the value of ENVSI of 0.39 which is also higher than the EEMD.

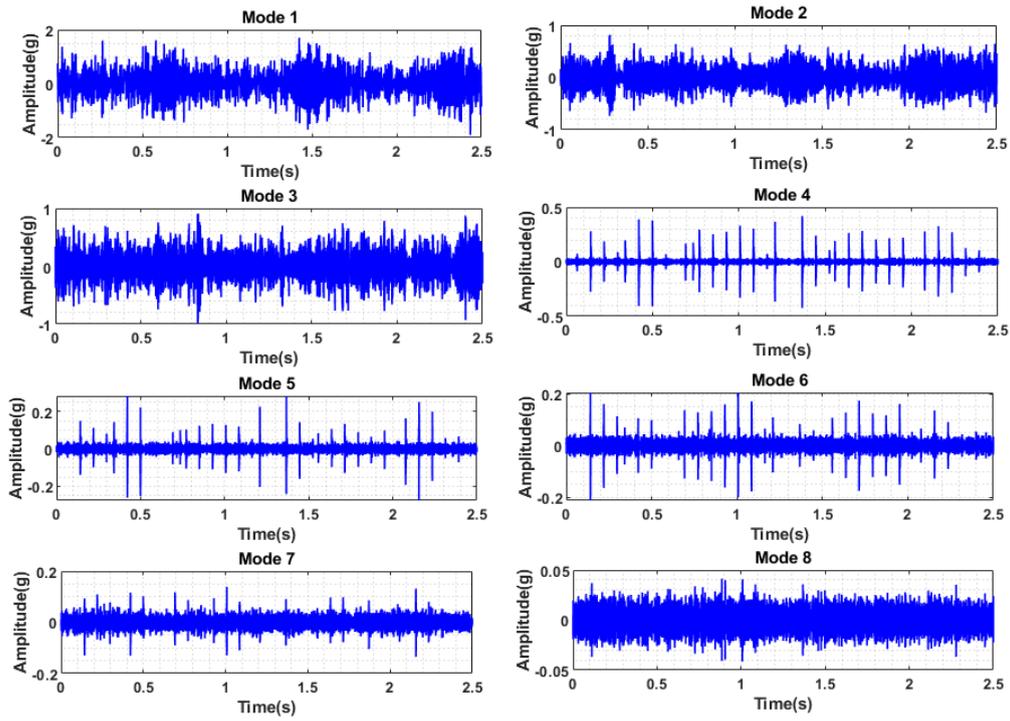

Fig. 18 Decomposition of vibration signal of bearing from pulley through VMD

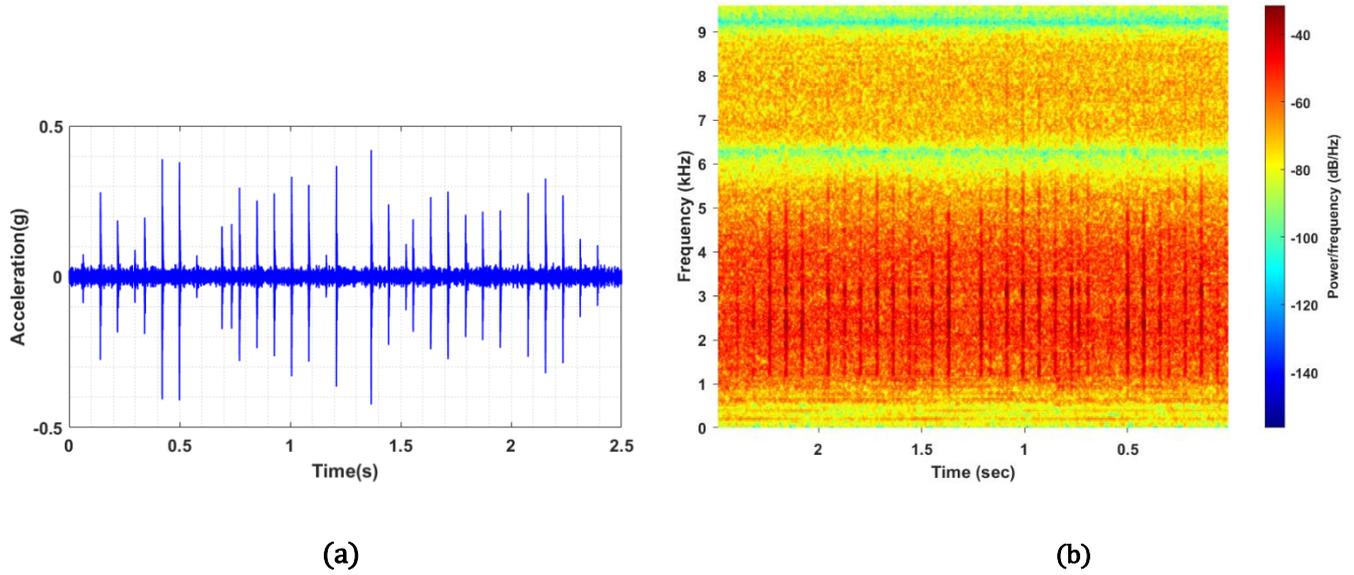

(a)  (b)

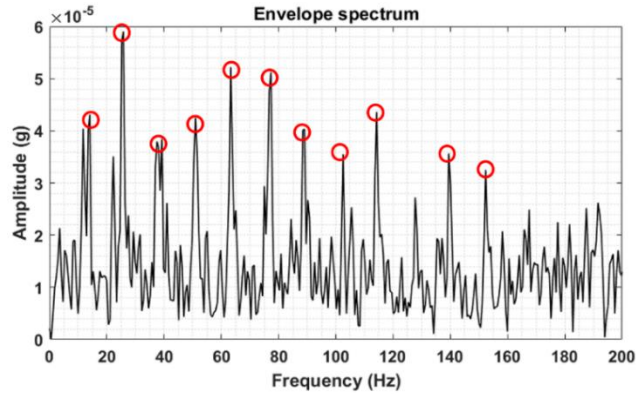

(c)

Fig. 19 Result of VMD-based decomposition of vibration signal from faulty bearing from pulley

In case of the acoustic signal of the defective bearing from the idler, 12 modes have been obtained by the VMD. The best mode obtained by the VMD is mode 4 as shoen in Fig. 21(a). The value of kurtosis is higher than EEMD but lower than SEEMD. The informative frequency band can be easily identified in the spectrogram obtained by the VMD. Whereas the SEEMD gives a more prominent informative frequency band than that of VMD as shown in Fig. 21 (b). The fault frequencies obtained by VMD are the same as those of SEEMD. However, the value of the ENVSI obtained by the SEEMD is 0.56 whereas the VMD gives the value of ENVSI of 0.48 which is also higher than the EEMD.

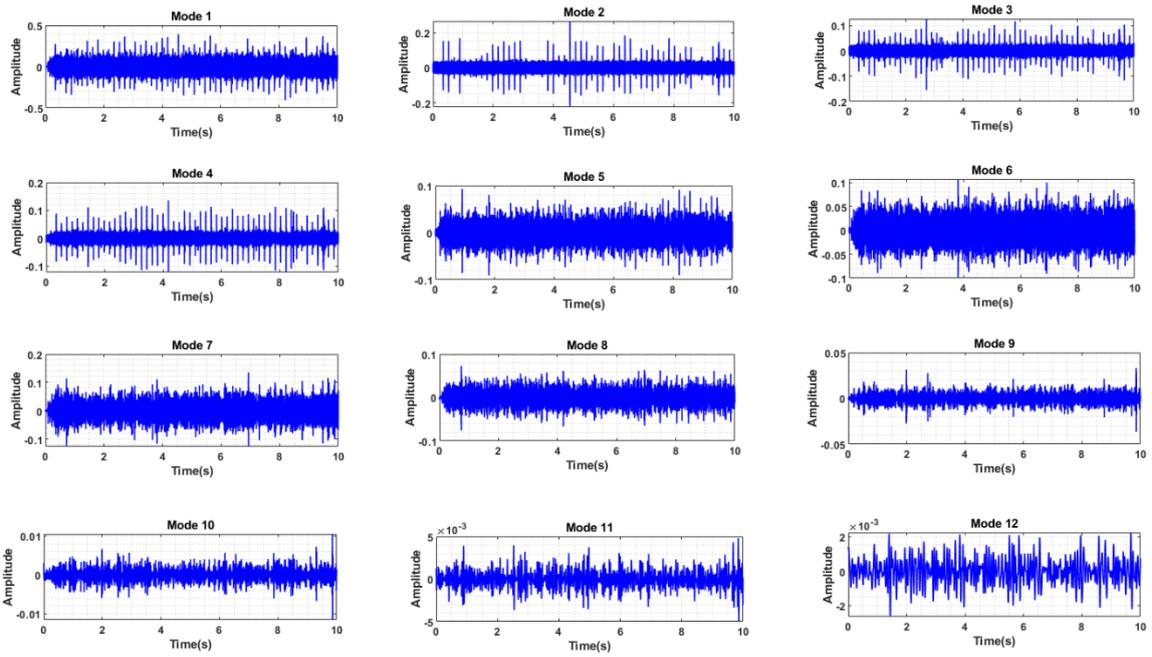

Fig. 20 Decomposition of acoustic signal of faulty bearing from idler through VMD

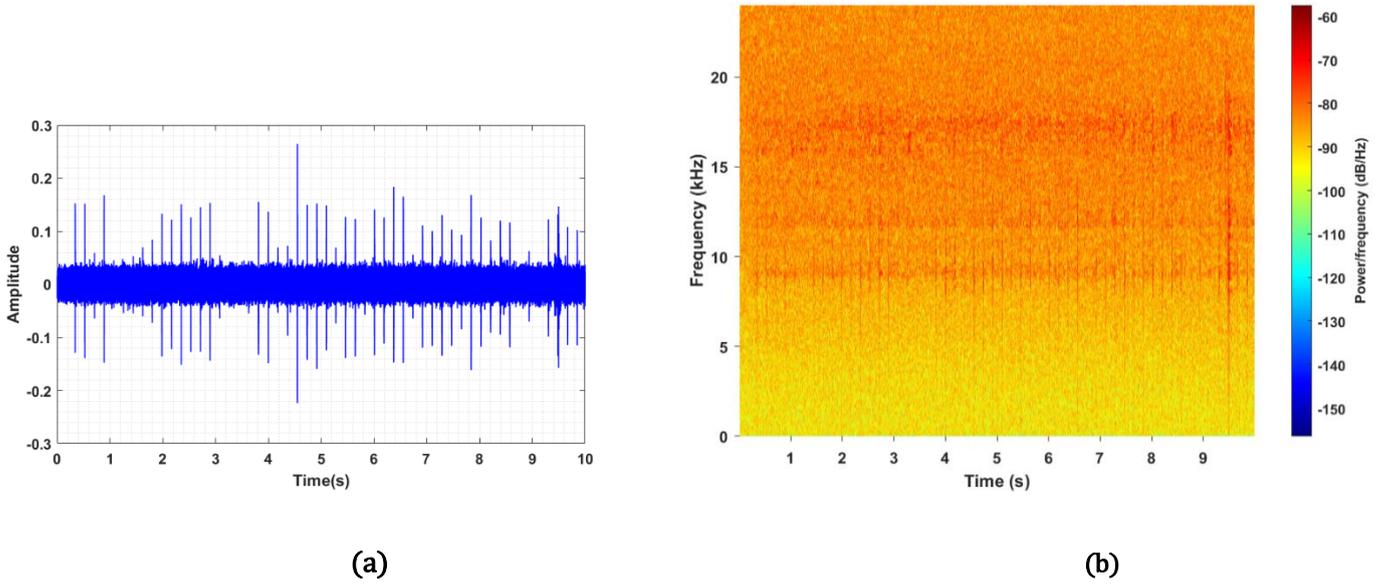

(a)                                                 (b)

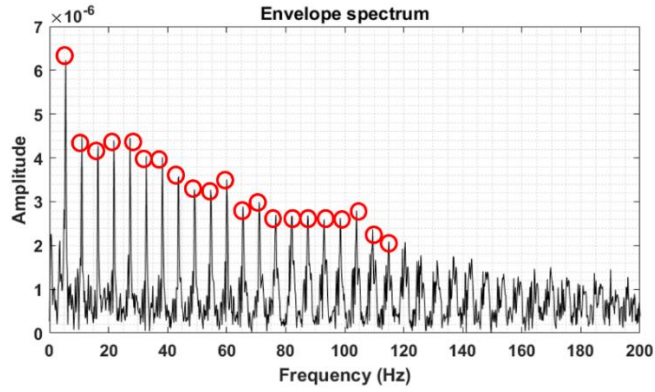

(c)

Fig. 21 Result of VMD-based decomposition of acoustic signal from faulty bearing used in idler

Also, the value of kurtosis both in vibration and an acoustic signal is higher for output signal based on SEEMD when compared to that of a EEMD and VMD. The value of the kurtosis in each case has also been tabulated in Table. 2.

Table 2
Comparison of kurtosis value under different methods

| Source | Raw signal | Output signal based on EEMD | Output signal based on VMD | Output signal based on proposed SEEMD |
|---|---|---|---|---|
| | | Kurtosis value | | |
| Vibration signal | 3.0978 | 50.9651 | 88.5962 | 96.5438 |
| Acoiustic signal | 3.6854 | 52.8743 | 75.5271 | 95.3508 |

## 8. Conclusions

In this research, a novel technique i.e. single ensemble empirical mode decomposition (SEEMD) is put forward to detect the local faults in rolling element bearings. The following conclusions have been drawn:

1. In SEEMD, initially, a fractional Gaussian noise (FGN) is added to the raw signal which is further multiplied by convoluted white Gaussian noise to obtain the different IMFs.
2. The proposed SEEMD does not require a number of ensembles from the addition or

subtraction of noise every time while processing the signals just like EEMD.

3. In SEEMD, only single shifting process is required when compared to EEMD which reduces the computation time significantly.
4. The proposed SEEMD proved to be beneficial while dealing with non-Gaussian or non-stationary perturbing signals.
5. SEEMD is also applied to the raw signals (both vibration and acoustic signals) obtained from the mining industries as these signals are difficult to analyze since since cyclic impulsive components are obscured by noise and other interference. The proposed SEEMD is better able to extract the SOI by reducing the ce interferences from other machinery components and the environment. which is further validated by spectrogram and envelope spectrum.
6. The SEEMD is compared with traditional methods such as EEMD and VMD based spectrogram, envelope spectrum and ENVSI. It has been proved that the proposed SEEMD is better able to remove the noises and other interferences.

## Acknowledgements

The work of RZ is supported by the National Center of Science under the Sheng2 project No. UMO- 2021/40/Q/ST8/00024 "NonGauMech - New methods of processing non-stationary signals (identification, segmentation, extraction, modeling) with non-Gaussian characteristics to monitor complex mechanical structures".